\documentclass[aps,prb,superscriptaddress,twocolumn,showpacs,amsmath,amssymb]{revtex4}

\usepackage{graphicx}

\begin{document}
\title{Superconductivity-Induced Self-Energy Evolution of the Nodal Electron in Optimally-Doped Bi2212}
\author{W. S. Lee}
\affiliation {Department of Physics, Applied Physics, and Stanford Synchrotron Radiation Laboratory, Stanford University, Stanford, CA 94305}
\author{W. Meevasana}
\affiliation {Department of Physics, Applied Physics, and Stanford Synchrotron Radiation Laboratory, Stanford University, Stanford, CA 94305}
\author{S. Johnston}
\affiliation {Department of Physics, Applied Physics, and Stanford Synchrotron Radiation
Laboratory, Stanford University, Stanford, CA 94305}
\affiliation {Department of Physics, University of Waterloo,Waterloo, Ontario, Canada N2L 3G1}
\author{D. H. Lu}
\affiliation {Department of Physics, Applied Physics, and Stanford Synchrotron Radiation Laboratory, Stanford University, Stanford, CA 94305}
\author{I. M. Vishik}
\affiliation {Department of Physics, Applied Physics, and Stanford Synchrotron Radiation Laboratory, Stanford University, Stanford, CA 94305}
\author{R. G. Moore}
\affiliation {Department of Physics, Applied Physics, and Stanford Synchrotron Radiation Laboratory, Stanford University, Stanford, CA 94305}
\author{H. Eisaki}
\affiliation {Nanoelectronic Research Institute, National Institute of Advanced Industrial Science and Technology, 1-1-1 central 2, Umezono, Tsukuba, Ibaraki, 305-8568, Japan}
\author{N. Kaneko}
\affiliation {Department of Physics, Applied Physics, and Stanford Synchrotron Radiation Laboratory, Stanford University, Stanford, CA 94305}

\author{T. P. Devereaux}
\affiliation {Department of Physics, Applied Physics, and Stanford Synchrotron Radiation Laboratory, Stanford University, Stanford, CA 94305}

\author{Z. X. Shen}
\affiliation {Department of Physics, Applied Physics, and Stanford Synchrotron Radiation Laboratory, Stanford University, Stanford, CA 94305}


\date{\today}

\begin{abstract}
The temperature dependent evolution of the renormalization effect in optimally-doped Bi2212 along the nodal direction has been studied via angle-resolved photoemission spectroscopy. Fine structure is observed in the real part of the self-energy (Re$\Sigma$), including a subkink and maximum, suggesting that electrons couple to a spectrum of bosonic modes, instead of just one mode. Upon cooling through the superconducting phase transition, the fine structures of the extracted Re$\Sigma$ exhibit a two-processes evolution demonstrating an interplay between kink renormalization and superconductivity. We show that this two-process evolution can be qualitatively explained by a simple Holstein model in which a spectrum of bosonic modes is considered.

\end{abstract}

\pacs{Valid PACS appear here}
\maketitle

An important issue concerning the microscopic mechanism of high-$T_c$ superconductivity is whether Cooper pairs are bound by collective modes, analogous to the role of phonons in conventional BCS superconductors. In conventional superconductors, the electron mass is renormalized by phonons,  causing fine structures in the single particle spectral function as observed in tunneling spectra (e.g. Ref. \raisebox{-1.2ex}{\Large\cite{BCS:tunneling}}). In the high-$T_c$ superconducting copper oxides, an electronic renormalization effect in the form of a kink in the band dispersion along the diagonal of the Brillouin zone (nodal direction), has been discovered by angle-resolved photoemission spectroscopy \cite{Mode:Pasha, group_review}, suggesting that the electrons couple to some bosonic mode(s). Although the identity of the bosonic mode, lattice \cite{Lanzara:phonon,Tanja:B1g:PRL,XJZhou:multiple_mode,non:doping_dependence_monolayer}  or magnetic \cite{Mode:Kaminski,mode:sato:BiFamily,A.A.Kordyuk:kink_with_self_consistent,Mode:Gromko}, has been the focus of great interest in connection  to pairing, aspects of this renormalization effect still remain unclear. In particular, the lack of a superconductivity-induced shift of the energy of the kink remains an important puzzle regardless of whether this bosonic mode is a phonon or a magnetic mode.

\begin{figure} [b]
\includegraphics [clip, height=2.0 in]{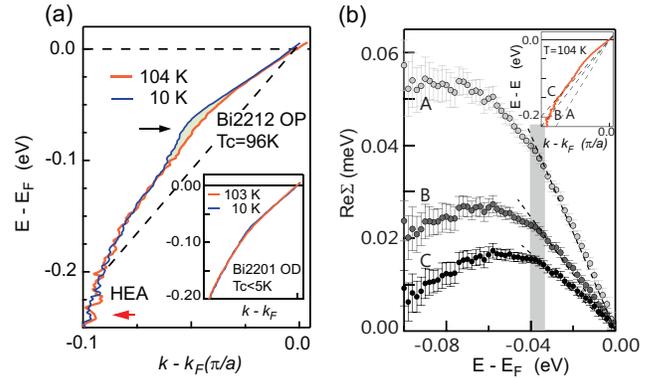}
\caption{\label{Fig1:puzzle_and_bare_band} (Color online) (a) The MDC-derived band dispersions of optimally-doped Bi2212 at temperature above and below $T_c$. The apparent dispersion kink and the onset of the ``high energy anomaly" (HEA), are approximately indicated by the longer and shorter arrows, respectively. The shaded area highlights the difference of band dispersion between these two temperatures. The dashed line is the ``bare band" for extracting the Re$\Sigma$ as described in the text. Inset shows the MDC-derived band dispersion of non-superconducting Bi2201 crystals measured at similar temperatures. (b) Re$\Sigma$ extracted using different choices of the bare band as plotted and labeled in the inset. The shaded area indicates the energy of the subkinks discussed in Fig. \ref{Fig2:SelfEnergy_high_and_Low_temperature}.}
\end{figure}

According to the theoretical calculations using an Eliashberg formalism, a coupling to a sharp bosonic mode in a $d$-wave superconductor with a maximum gap $\Delta_0$ yields a dispersion kink along the nodal direction at an energy of $\Omega + \Delta_0$ \cite{Norman:Theoory:SpinResonance,SandvikAndDoug:Theory,Tom:B1g:PRL}. Above $T_c$, the kink position is expected to shift to a lower binding energy due to the closing of the superconducting gap; thus, a different kink position than that at temperatures below $T_c$ is expected. This, however, has not been observed; in fact, the ``apparent" kink position in the nodal dispersion remains at approximately the same energy ($\sim$ 70 meV) in both the normal and superconducting states \cite{Mode:Kaminski,Mode:Gromko,mode:sato:BiFamily}. This lack of energy shift is a long standing puzzle \cite{Kulic:Absent_of_T_dep_Kink:Theory}, which seems to contradict theoretical predictions, casting doubts on the interpretation of the dispersion kink as a renormalization effect due to electron-boson coupling \cite{BYCZUK:Kink_Strong_Correlated:Theory}.

Here, we focus on the band dispersion along (0,0)-($\pi$,$\pi$) direction (nodal direction) and report temperature dependent measurements on optimally-doped Bi-2212. Compared to other cuprates with lower $T_c$, the large superconducting gap ($\Delta_0 \sim$ 40 meV) of optimally-doped Bi-2212 provides a better opportunity to resolve the superconductivity-induced phenomena in the nodal electron renormalization. A reference experiment on heavily overdoped non-superconducting Bi2201 has also been performed as a comparison.

High quality single crystals of optimally doped Bi$_2$Sr$_2$Ca$_{0.92}$Y$_{0.08}$Cu$_2$O$_{8+\delta}$ (Bi2212 OP, $T_c$ = 96K) \cite{Bi2212:inhomogeneity:Hiroshi} were selected to perform the temperature dependence experiments. For the reference experiment, single crystals of heavily overdoped non-superconducting Pb$_{0.38}$Bi$_{1.74}$Sr$_{1.88}$CuO$_{6+\delta}$ (Bi2201 OD, $T_c<$ 5K) were also measured in the same temperature range. The main results of Bi2212 were collected by using He I light (21.2 eV) from a monochromated and modified Gammadata He Lamp with a Scienta-2002 analyzer. The data on Bi2201 were collected by using 19 eV photons at Stanford Synchrotron Radiation Laboratory (SSRL) beamline 5-4 with Scienta-200 analyzer. The energy resolution is $\sim$10 meV and angular resolution $\sim$0.35$^o$. The sample is cleaved and measured in ultra high  vacuum ($ < 4\times10^{-11}$ Torr.) to maintain a clean surface. The same temperature dependence experiments on Bi2212 were repeated at SSRL, which reproduces the results obtained from the He lamp.

The band dispersion along the nodal direction is extracted in the usual manner by fitting the momentum distribution curve (MDC) to a Lorentizan function \cite{Mode:Kaminski}. The band dispersions from above and below $T_c$ are shown in Fig. \ref{Fig1:puzzle_and_bare_band}(a). As can be seen, the apparent kink position in the dispersion at both temperatures is located at about 70 meV presenting the aforementioned puzzle . Upon close examination, there is a difference between the dispersions above and below $T_c$  around the apparent kink position (shaded area in Fig. \ref{Fig1:puzzle_and_bare_band} (a)). Although this difference is subtle, it is significantly larger than that in non-superconducting Bi2201 (inset), where the difference is due to thermal broadening.  This observation implies that the dispersion difference seen in the case of optimally doped Bi2212 is a superconductivity-induced phenomena, rather than a trivial thermal broadening effect. We note that these observations are consistent with the previously published data \cite{Mode:Kaminski,Mode:Gromko,mode:sato:BiFamily} but the puzzle of missing energy shift of the dispersion kink across the superconducting phase transition has not yet been addressed.

This subtle difference could be further analyzed by extracting the real part of the self-energy Re$\Sigma$, obtained by subtracting the MDC-derived dispersion from an assumed bare band. A rigorous determination of the bare band for this low energy renormalization effect (the kink structures) is generally difficult, and further complicated by the recently discovered high energy anomaly HEA\cite{HEA:non}. Nonetheless, to zeroth order, the low energy renormalization effect could be assumed to manifests itself on an \emph{effective} bare band which accounts for all renormalization effects from the HEA and other possible contributions. We assume this effective bare band can be approximated by a linear dispersion connected from the Fermi surface crossing to 200 meV below $E_F$ on the dispersion before the onset of the HEA, as illustrated by the dashed line in Fig. \ref{Fig1:puzzle_and_bare_band}(a). We also argue that the temperature dependence of this bare band is small and negligible in the temperature range of interest, since no significant temperature dependence of the band dispersion is seen at energies greater than 150 meV below $E_F$ (Fig. \ref{Fig1:puzzle_and_bare_band}(a)).

\begin{figure}[t]
\includegraphics [clip, width=3.25 in]{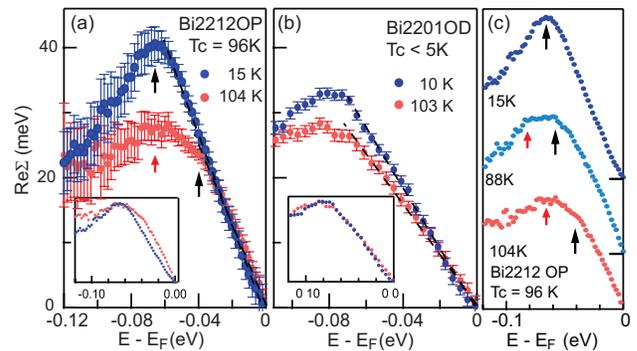}
\centering
\caption{\label{Fig2:SelfEnergy_high_and_Low_temperature} (Color online) Re$\Sigma$ in the normal state and superconducting state for: (a) optimally-doped Bi2212, (b) Non-superconducting heavily overdoped Bi2201. Insets plot Re$\Sigma$ with their maximum normalized to illustrate the difference of the shapes. The arrows indicate the energy positions of the kink and subkink. The errors bars were estimated from the 99.7\% confidence interval of the fitted MDC peak positions. (c) The evolution of Re$\Sigma$ across the superconducting phase transition.}
\end{figure}

The extracted Re$\Sigma$ of optimally-doped Bi2212 above and below $T_c$ are plotted in Fig. \ref{Fig2:SelfEnergy_high_and_Low_temperature}(a). At both temperatures, the maximum of Re$\Sigma$, essentially the \emph{apparent} kink position in the band dispersions, is located at approximately the same energy, $\sim$70 meV. This coincidence of the maximum positions verifies previous observations of the absent shift of the apparent kink position. Importantly, an additional feature can now be visualized in the extracted Re$\Sigma$ from the data taken above $T_c$: a shoulder or ``subkink" at about $\sim 35\pm 4$ meV where the slope of Re$\Sigma$ changes abruptly. We stress that the observation of a subkink in Re$\Sigma$ above $T_c$ in optimally doped Bi2212 system is a robust experimental feature; it is not only well reproducible in different sets of our data, but also has been seen in the published data, where this subkink feature was not discussed \cite{Mode:Gromko,T_Yamasaki:Hiroshima:Bi2212}. We also note that while the magnitude of the overall self-energy is dependent on the choice of the bare band, the position of this subkink is relatively immune to this choice as illustrated in Fig. \ref{Fig1:puzzle_and_bare_band}(b). Notably, the fine structures in Re$\Sigma$ have also been observed in other families of cuprates in the normal state, which has been interpreted as the electrons coupled to a spectrum of phonons \cite{XJZhou:multiple_mode, non:doping_dependence_monolayer}. More interestingly, as further shown in Fig. \ref{Fig2:SelfEnergy_high_and_Low_temperature} (a), comparison of Re$\Sigma$ at different temperatures reveals a clear change of its shape across the superconducting phase transition; at a temperature well below $T_c$, only a maximum with no ``subkink" in Re$\Sigma$ (dark markers) can be identified. This temperature dependent shape change is further emphasized in a plot of the normalized Re$\Sigma$ (inset).

Further support could be drawn from a reference experiment on heavily overdoped non-superconducting Bi2201 crystals, whose Re$\Sigma$ at two temperatures are plotted in Fig. \ref{Fig2:SelfEnergy_high_and_Low_temperature}(b). Compared to the data of optimally-doped Bi2212 (Fig. \ref{Fig2:SelfEnergy_high_and_Low_temperature}(a)), there are two major differences: (1) no detectable shape change of Re$\Sigma$ when the temperature varies from 103 K to 10 K; the only change here is a slight difference of the magnitude of the maximum due the thermal broadening effect. (see also the normalized Re$\Sigma$ plotted in the inset). (2) No prominent subkink structure is seen. These two differences strongly suggest that the thermal broadening effect can not explain the shape of the Re$\Sigma$ and its temperature dependent evolution observed in optimally-doped Bi2212; otherwise, a similar behavior of Re$\Sigma$ should have been observed in this reference experiment. Therefore, what has been observed in optimally-doped Bi2212 is a superconductivity-induced phenomena, and the subkink in the Re$\Sigma$ in the normal state is a generic feature of optimally-doped Bi2212. The absence of a subkink in heavily overdoped Bi2201 is related to a material and doping dependent coupling between the electrons and the spectrum of bosonic modes, which has been discussed elsewhere \cite{non:doping_dependence_monolayer,XJZhou:multiple_mode}.

\begin{figure}[t]
\includegraphics [clip, width=3.25 in]{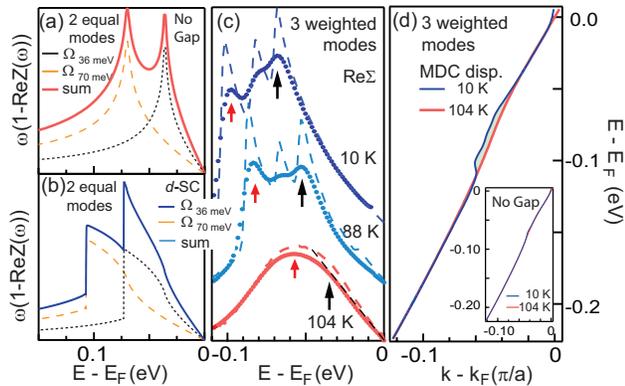}
\centering
\caption{\label{Fig3:Simulations_and detailed_T_dep} (Color online) (a-b) The self-energy parameter $Z(\omega)$ of 2 equally strong Einstein modes in the case of (a) normal state and (b) a $d-$wave superconductor with $\Delta_0$ =37 meV. (c) Extracted Re$\Sigma$ by applying MDC-analysis to the spectral functions (simulated ARPES data) calculated with 3 weighted Einstein modes in a tight-bind band structure at different temperatures from normal state (104K) to superconducting state (88K with $\Delta_0$ = 20 meV, and 10 K with $\Delta_0$ = 37 meV). The arrows indicate the energy positions of the kink and subkink. (d) Band-dispersion along the nodal direction extracted from the spectral functions calculated by the same condition described in (c).  The shaded area highlights the difference of band dispersion between these two temperatures. Inset plots the MDC-dispersions at 104 K and 10 K extracted for the case of a non-superconductor.}
\end{figure}

The evolution of extracted Re$\Sigma$ from above $T_c$ to below $T_c$ in optimally-doped Bi2212 is shown in Fig. \ref{Fig2:SelfEnergy_high_and_Low_temperature}(c), where the data slightly below $T_c$ at 88K provides important information not available in earlier experiments. Two process are involved. First, both the maximum (shorter arrow) and subkink (longer arrow), shift toward higher binding energy due to the opening of a superconducting gap. This energy shift is expected in the Eliashberg formalism lending further support for interpreting fine structures (subkink and maximum) in Re$\Sigma$ as coupling to a spectrum of bosonic modes.   Second,  we also observed a non-trivial redistribution of the relative weight between the lower and higher energy renormalization features across the superconducting transition. The lower energy feature (longer arrow) gradually dominates in Re$\Sigma$, such that  the slope between the two arrows gradually changes from negative to positive; at a temperature well below $T_c$ (15 K), the lower energy feature dominates and sets the maximum of Re$\Sigma$. The higher energy feature is then irresolvable. This gives an impression that there is a superconducting-induced change of mode-coupling, favoring the lower energy mode at the expense of the higher one as the temperature crosses $T_c$.

To qualitatively understand this spectral weight redistribution of renormalization features in Re$\Sigma$, we consider a model with a simple bosonic spectrum consisting of only two Einstein modes with energies $\Omega$ = 36 and 70 meV, respectively. The mode energies were chosen to approximately match the energy of the subkink and maximum of our Re$\Sigma$ data. The electronic self energy $\omega(1-Z(\omega))$ is plotted in Figs. \ref{Fig3:Simulations_and detailed_T_dep} (a) and (b) for the case of the normal state (i.e. ``no gap") and a $d$-wave superconductor with $\Delta_0$ = 37 meV, respectively \cite{E_P_interaction_in_metal:Grimvall,SandvikAndDoug:Theory}. Here we take the temperature $T$=0 and assume a circular Fermi surface. In both figures the coupling strength of each mode to the electrons is identical ($\Omega \cdot \lambda$ = 52 meV is assumed). Working in weak coupling, the total self-energy is a sum of the contributions of both modes as shown in the figures.

When the superconducting gap is absent, the two peaks of the total self-energy, renormalization features caused by the two coupled modes, are equally strong. When the superconducting gap is present, the lower energy renormalization feature appears to be stronger than the other, even though the coupling strength of these two modes are identical. The enhancement of the lower energy renormalization feature is essentially what has been observed experimentally (Figs. \ref{Fig2:SelfEnergy_high_and_Low_temperature} (a) and (c)). According to our model, this enhancement is primary due to the highly asymmetric shape of $\omega(1-Z(\omega))$ near the singularity for each mode in the superconducting state; it increases monotonically from $E_F$ up to the singularity $\Omega_{mode}+\Delta_0$, then drops suddenly to a value close to zero (Fig. \ref{Fig3:Simulations_and detailed_T_dep}(b)). Thus, when summing self-energies of the two modes for the total contribution, the self-energy peak induced by the lower energy mode ``sits" on the rising slope of the self-energy of the higher energy mode; as a consequence, it appears to be enhanced; while for the higher energy mode, its peak can only sit on the flat tail of the self-energy of the lower energy mode, hence, no significant enhancement for the higher energy feature in the total self-energy. This is different than the ``no gap" case, where the shape of the self-energy is more or less symmetric with respect to the singularity (logarithmic divergence, Fig. \ref{Fig3:Simulations_and detailed_T_dep}(a)); thus, the higher energy renormalization feature also acquires a similar enhancement from the tail of the lower renormalization feature.

The key features demonstrated in Figs. \ref{Fig3:Simulations_and detailed_T_dep}(a) and (b) remain valid even if the bosonic spectrum consists of more than two modes in a generic cuprate band structure at finite temperatures. In Fig. \ref{Fig3:Simulations_and detailed_T_dep}(c-d), we present the simulated renormalization effect at finite temperatures, using the band structure of Ref. \raisebox{-1.2ex}{\Large\cite{Norman:Theoory:SpinResonance}}, resulting from coupling to three modes at energies 36, 50 and 70 meV, corresponding to energies of typical optical phonons in cuprates. The relative strength of each mode coupling has been adjusted such that the shape of the Re$\Sigma$ in the normal state (104 K)  looks similar to our 104 K  data in Fig. \ref{Fig2:SelfEnergy_high_and_Low_temperature} (a) \cite{coupling:note}; the coupling strength then remains unchanged for the superconducting state calculations ($d$-wave gap, at 88K with $\Delta_0$=20 meV,  and 10K with $\Delta_0$=37 meV). The spectral functions along the nodal direction were calculated and then convoluted with a 2D Gaussian response function with 10 meV in energy and 0.012 $\pi/a$ in momentum direction to simulate finite instrument resolution effect. The MDC analysis used for actual ARPES data is applied to this simulated spectral functions to extract the renormalized band dispersions. As shown in Fig. \ref{Fig3:Simulations_and detailed_T_dep} (d), the difference between the MDC-dispersions at these two temperatures (shaded area) is significantly larger than that caused by thermal broadening (the inset). This reproduces our experimental observations on optimally-doped Bi2212 and the reference experiment on non-superconducting heavily overdoped Bi2201 (see  Fig. \ref{Fig1:puzzle_and_bare_band} (a)).

The temperature dependent evolution of the self-energy extracted from MDC dispersion is shown in Fig. \ref{Fig3:Simulations_and detailed_T_dep}(c). As expected, the ``peaks" in the Re$\Sigma$ corresponding to each individual mode is less resolvable due to the superposition with others and the smearing by the resolution effects, leaving only subkinks in the Re$\Sigma$, reminiscent of the data shown in Fig. \ref{Fig2:SelfEnergy_high_and_Low_temperature} and the extracted Re$\Sigma$ in other cuprates \cite {XJZhou:multiple_mode,non:doping_dependence_monolayer}. Our simulations also demonstrate that the multimode nature can still survive at a temperature of 104 K in the form of self-energy shape after convolving with instrument resolution. Furthermore, the two-process evolution of the Re$\Sigma$ observed in our data (Fig. \ref{Fig2:SelfEnergy_high_and_Low_temperature} (c)) is also qualitatively reproduced. We emphasize that the \emph{apparent} kink (the broad maximum of Re$\Sigma$) in the band dispersion at 15 K and 104 K is caused by different bosonic modes, \emph{not} by the same mode. On the contrary, in the case of non-superconducting Bi2201, the ``apparent" kink in the dispersion is caused by the same mode, since no temperature dependent energy shift of Re$\Sigma$ is observed (Fig. \ref{Fig2:SelfEnergy_high_and_Low_temperature}(b)). These results also warn that the ``apparent" kink alone does not tell a complete story of the renormalization effect in cuprates.

SSRL is operated by the DOE Office of Basic Energy Science, Division of Chemical Science and Material Science. This work is supported by DOE Office of Science, Division of Materials Science, with contract DE-FG03-01ER45929-A001 and NSF grant DMR-0604701.


\begin{thebibliography}{99}
\bibitem{BCS:tunneling}
J. M. Rowell, P. W. Anderson, and D. E. Thomas, Phys. Rev. Lett. \textbf{10}, 334 (1963); D. J. Scalapino, J. R. Schrieffer, and J. W. Wilkins, Phys. Rev. \textbf{148}, 263 (1966).

\bibitem{Mode:Pasha}
P. V. Bogdanov, A. Lanzara, S. A. Kellar, X. J. Zhou, E. D. Lu, W. J. Zheng, G. Gu, J.-I. Shimoyama, K. Kishio, H. Ikeda, R. Yoshizaki, Z. Hussain, and Z. X. Shen, Phys. Rev. Lett. \textbf{85}, 2581 (2000).

\bibitem{group_review}
A. Damascelli, Z.-X. Shen, and Z. Hussain, Rev. Mod. Phys. \textbf{75}, 473 (2003), and the references therein.

\bibitem{Lanzara:phonon}
A. Lanzara, P. V. Bogdanov, X. J. Zhou, S. A. Kellar, D. L. Feng, E. D. Lu, T. Yoshida, H. Eisaki, A. Fujimori, K. Kishio, J.-I. Shimoyama, T. Noda, S. Uchida, Z. Hussain, and Z.-X. Shen, Nature (London) \textbf{412}, 510 (2001).

\bibitem{Tanja:B1g:PRL}
T. Cuk, F. Baumberger, D. H. Lu, N. Ingle, X. J. Zhou, H. Eisaki, N. Kaneko, Z. Hussain, T. P. Devereaux, N. Nagaosa, and Z.-X. Shen, Phys. Rev. Lett. \textbf{93}, 117003 (2004).

\bibitem{XJZhou:multiple_mode}
 X. J. Zhou, Junren Shi, T. Yoshida, T. Cuk, W. L. Yang, V. Brouet, J. Nakamura, N. Mannella, Seiki Komiya, Yoichi Ando, F. Zhou, W. X. Ti, J. W. Xiong, Z. X. Zhao, T. Sasagawa, T. Kakeshita, H. Eisaki, S. Uchida, A. Fujimori, Zhenyu Zhang, E. W. Plummer, R. B. Laughlin, Z. Hussain, and Z.-X. Shen, Phys. Rev. Lett. \textbf{95}, 117001 (2005).

\bibitem{non:doping_dependence_monolayer}
W. Meevasana, N. J. C. Ingle, D. H. Lu, J. R. Shi, F. Baumberger, K. M. Shen, W. S. Lee, T. Cuk, H. Eisaki, T. P. Devereaux, N. Nagaosa, J. Zaanen, and Z.-X. Shen \emph{et al.}, Phys. Rev. Lett. \textbf{96}, 157003 (2006).

\bibitem{Mode:Kaminski}
A. Kaminski, M. Randeria, J. C. Campuzano, M. R. Norman, H. Fretwell, J. Mesot, T. Sato, T. Takahashi, and K. Kadowaki, Phys. Rev. Lett. \textbf{86}, 1070 (2001).

\bibitem{mode:sato:BiFamily}
T. Sato, H. Matsui, T. Takahashi, H. Ding, H.-B. Yang, S.-C. Wang, T. Fujii, T. Watanabe, A. Matsuda, T. Terashima, and K. Kadowaki, Phys. Rev. Lett. \textbf{91}, 157003 (2003).

\bibitem{A.A.Kordyuk:kink_with_self_consistent}
A. A. Kordyuk, S. V. Borisenko, V. B. Zabolotnyy, J. Geck, M. Knupfer, J. Fink, B. Bu"chner, C. T. Lin, B. Keimer, H. Berger, A. V. Pan, Seiki Komiya, and Yoichi Ando, Phys. Rev. Lett. \textbf{97}, 17002(2006).

\bibitem{Mode:Gromko}
A. D. Gromko, A. V. Fedorov, Y.-D. Chuang, J. D. Koralek, Y. Aiura, Y. Yamaguchi, K. Oka, Yoichi Ando, and D. S. Dessau, Phys. Rev. B \textbf{68}, 174520 (2003).

\bibitem{Norman:Theoory:SpinResonance}
M. Eschrig and M. R. Norman, Phys. Rev. B \textbf{67}, 144503 (2003).

\bibitem{SandvikAndDoug:Theory}
A. W. Sandvik, D. J. Scalapino, and N. E. Bickers, Phys. Rev. B \textbf{69}, 094523(2004).

\bibitem{Tom:B1g:PRL}
T. P. Devereaux, T. Cuk, Z.-X. Shen, and N. Nagaosa, Phys. Rev. Lett. \textbf{93}, 117004 (2004).

\bibitem{Kulic:Absent_of_T_dep_Kink:Theory}
M. L. Kulic and O. V. Dolgov, Phys. Rev. B \textbf{71}, 092505 (2005)

\bibitem{BYCZUK:Kink_Strong_Correlated:Theory}
K. Byczuk, M. Kollar, K. Held, Y.-F. Yang, I. A. Nekrasov, Th. Pruschke, and  D. Vollhardt, Nature Physics \textbf{3}, 168 (2007).

\bibitem{Bi2212:inhomogeneity:Hiroshi}
H. Eisaki, N. Kaneko, D. L. Feng, A. Damascelli, P. K. Mang, K. M. Shen, Z.-X. Shen, and M. Greven, Phys. Rev. B \textbf{69}, 064512 (2004).

\bibitem{HEA:non}
J. Graf, G.-H. Gweon, K. McElroy, S. Y. Zhou, C. Jozwiak, E. Rotenberg, A. Bill, T. Sasagawa, H. Eisaki, S. Uchida, H. Takagi, D.-H. Lee, and A. Lanzara, Phys. Rev. Lett. \textbf{98}, 067004 (2007);  W. Meevasana, X. J. Zhou, S. Sahrakorpi, W. S. Lee, W. L. Yang, K. Tanaka, N. Mannella, T. Yoshida, D. H. Lu, Y. L. Chen, R. H. He, Hsin Lin, S. Komiya, Y. Ando, F. Zhou, W. X. Ti, J. W. Xiong, Z. X. Zhao, T. Sasagawa, T. Kakeshita, K. Fujita, S. Uchida, H. Eisaki, A. Fujimori, Z. Hussain, R. S. Markiewicz, A. Bansil, N. Nagaosa, J. Zaanen, T. P. Devereaux, and Z.-X. Shen, Phys. Rev. B \textbf{75}, 174506 (2007).

\bibitem{T_Yamasaki:Hiroshima:Bi2212}
T. Yamasaki, K. Yamazaki, A. Ino, M. Arita, H. Namatame, M. Taniguchi, A. Fujimori, Z.-X. Shen, M. Ishikado, and S. Uchida Phys. Rev. B \textbf{75}, 140513(R)(2007).


\bibitem{E_P_interaction_in_metal:Grimvall}
G. Grimvall, \emph{The electron-phonon interaction in metals} (North-Holland, New York, 1981), p102.

\bibitem{coupling:note}
The scattering vertex ratio is $g_{36meV}:g_{50meV}:g_{70meV} \sim 3:2.24:2$ with total zone-averaged dimensionless coupling strength $\lambda \sim 0.3$.

\end{thebibliography}
\end{document}